\documentclass[twocolumn, superscriptaddress, secnumarabic, amssymb, nobibnotes, aps, prd, preprintnumbers, nofootinbib]{revtex4-1}

\input{header}

\begin{document}
   
\preprint{MI-HET-848}

\title{{\Large Neutrino-Portal Dark Matter Detection Prospects \\at a Future Muon Collider}}

\author{Jyotismita Adhikary}
\email{jyotismita.Adhikary@ncbj.gov.pl}
\affiliation{National Centre for Nuclear Research, Pasteura 7, Warsaw, PL-02-093, Poland}

\author{Kevin J. Kelly}
\email{kjkelly@tamu.edu}
\affiliation{Department of Physics and Astronomy, Mitchell Institute for Fundamental Physics and Astronomy, Texas A\&M University, College Station, TX 77843, USA}

\author{Felix Kling}
\email{felix.kling@desy.de}
\affiliation{Deutsches Elektronen-Synchrotron DESY, Notkestr.~85, 22607 Hamburg, Germany}

\author{Sebastian Trojanowski}
\email{sebastian.Trojanowski@ncbj.gov.pl}
\affiliation{National Centre for Nuclear Research, Pasteura 7, Warsaw, PL-02-093, Poland}

\begin{abstract}
With no concrete evidence for non-gravitational interactions of dark matter to date, it is natural to wonder whether dark matter couples predominantly to the Standard Model (SM)'s neutrinos. 
Neutrino interactions (and the possible existence of additional neutrinophilic mediators) are substantially less understood than those of other SM particles, yet this picture will change dramatically in the coming decades with new neutrino sources. 
One potential new source arises with the construction of a high-energy muon collider (MuCol) -- due to muons' instability, a MuCol is a source of high-energy collimated neutrinos.
Importantly, since the physics of muon decays (into neutrinos) is very well-understood, this leads to a neutrino flux with systematic uncertainties far smaller than fluxes from conventional high-energy (proton-sourced) neutrino beams.
In this work, we study the capabilities of a potential neutrino detector, ``MuCol$\nu$,'' placed ${\sim}$100~m downstream of the MuCol interaction point.
The MuCol$\nu$ detector would be especially capable of searching for a neutrinophilic mediator $\phi$ through the mono-neutrino scattering process $\nu_\mu N \to \mu^+ \phi X$, exceeding searches from other terrestrial approaches for $m_\phi$ in the ${\sim}$few MeV -- ten GeV range. 
Even with a $10$~kg-yr exposure, MuCol$\nu$ is capable of searching for well-motivated classes of thermal freeze-out and freeze-in neutrino-portal dark matter.
\end{abstract}

\maketitle 

\section{Introduction}
\label{intro}

Neutrinos and their interactions reside among the largest open questions in fundamental physics today.  The Standard Model (SM) of particle physics contains no definite mechanism by which neutrinos acquire mass which has driven immense experimental and theoretical activity since the observation of neutrino oscillations, a phenomenon that requires massive neutrinos. Simultaneous experimental and theoretical activity has scrutinized the completeness of the SM as a quantum field theory describing nature; if new physics exists between the scales of electroweak symmetry breaking and the Planck scale, it has yet to be discovered in our highest-energy facility, the Large Hadron Collider (LHC). One final mystery that remains unaddressed by the SM is the overwhelming evidence for dark matter (DM) in the universe; if it has interactions with the SM beyond gravitational ones, these are yet undetermined.

As experimental particle physics evolves, new strategies and approaches are continuously being developed to better understand fundamental particles and their interactions.
One promising approach to address many of the aforementioned puzzles in the future is the construction of a muon collider (MuCol), capable of operation at several-TeV energies~\cite{InternationalMuonCollider:2024jyv, MuCoL:2024oxj}.
While a great deal of physics can be explored by observing such high-energy muon-antimuon collisions, in this work, we focus on another aspect of such facilities: the unavoidable source of collimated, high-energy neutrinos produced as a byproduct.
Even with enormous Lorentz-boost factors, stored muons will decay quickly enough in these beamlines, leading to a well-characterized flux of neutrinos.
By deploying a relatively small neutrino detector nearby, yet aligned with the muon beam, we have the opportunity to study the interactions of TeV-scale neutrinos with a well-understood flux, allowing for great sensitivity to new physics. Hereafter, we refer to this detector as MuCol$\nu$.

One intriguing possibility that has garnered attention recently is that neutrinos, via some new mediator particle, carry the strongest interaction with dark matter particles.
This situation has gained visibility for two major reasons: from a model-building perspective, it is not too difficult to postulate a neutrophilic mediator~\cite{Berryman:2022hds} in the ${\sim}$MeV-GeV mass range that couples strongly to neutrinos and dark matter, but not (at least not strongly) to charged SM fermions.
Second, since neutrino interactions are the poorest-known among SM particles, there is ample possibility for novel interactions to exist.
In this paper, we explore the possibility of detecting such a neutrinophilic mediator in the MuCol$\nu$ detector close by to a muon collider's storage ring, using the potential for a connection to dark matter as additional motivation.

\begin{figure*}[thb]
\centering
\includegraphics[width=0.48\textwidth]{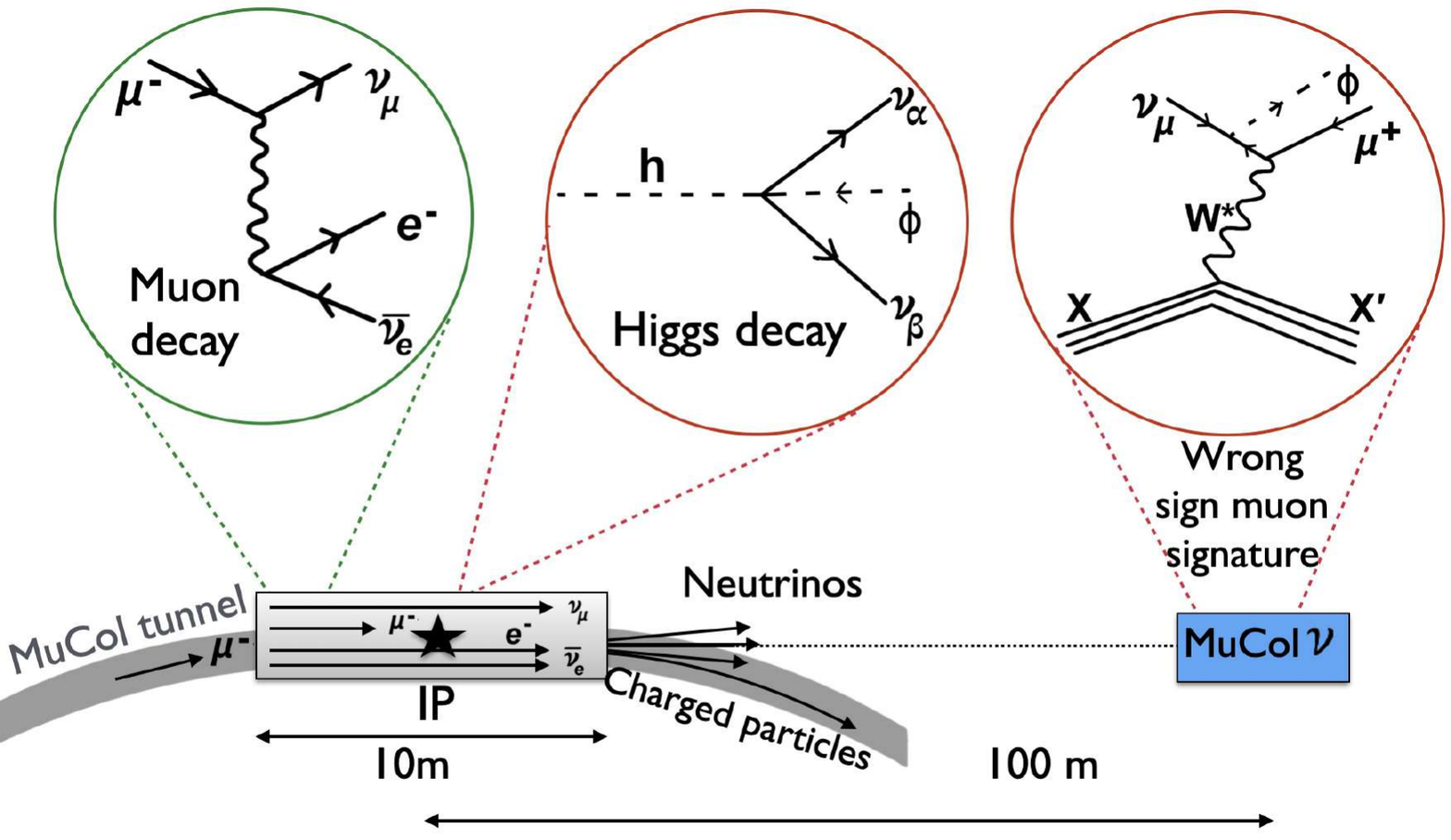}
\includegraphics[width=0.48\textwidth]{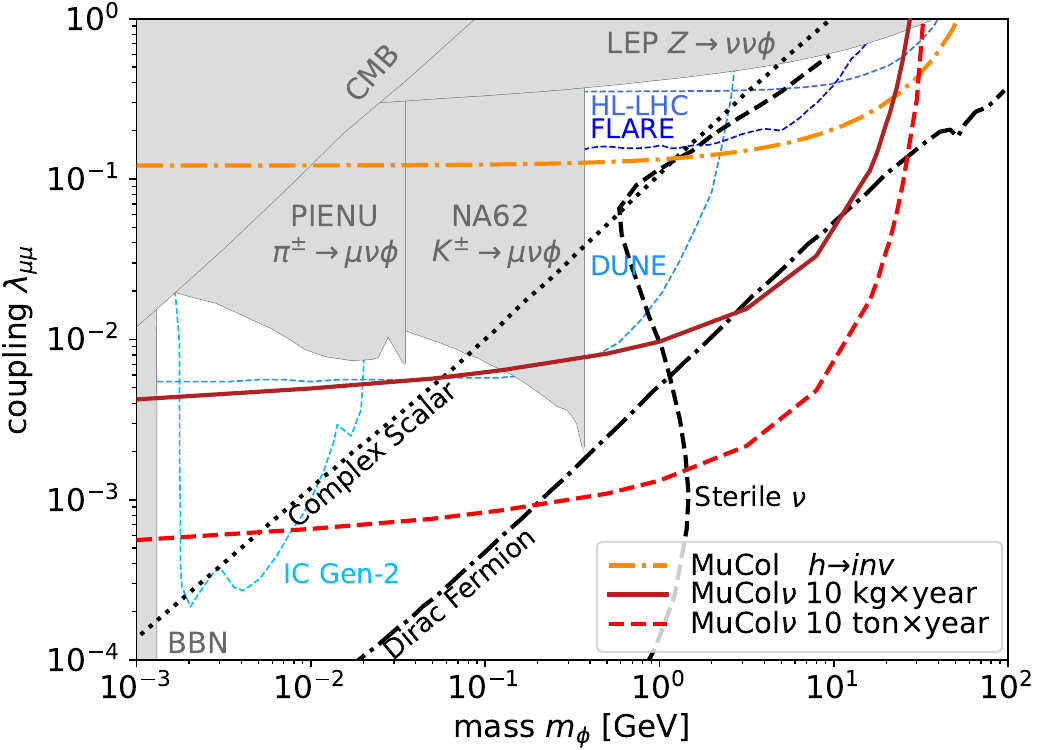}
\caption{\textbf{Left: Schematic diagram of the main focus of this paper.} In the left bubble, we show how stored muons in a muon collider (MuCol) tunnel decay into SM neutrinos. Downstream (right bubble), a hypothetical neutrino detector can search for mono-neutrino scattering with a wrong-sign muon due to emission of a lepton-number-charged mediator. Simultaneously (center), precision measurement of Higgs decays in the MuCol detectors can further constrain the new-physics scenario of interest. \textbf{Right: Landscape of the neutrinophilic mediator.} We show the sensitivity of neutrino scattering measurements using a 10~kg and 10~ton neutrino detector as well as precision measurements of the invisible decay width of the Higgs on the parameter space of the neutriniphilic mediator. Existing constraints are shown as gray shaded area while the projected sensitivity of other proposed searches is shown as blue dashed lines. See text for details.}
\label{fig:schematic}
\end{figure*}

The main idea of this work is summarized in the left panel of \cref{fig:schematic}. 
Muons in the straight parts of the beam pipe close to the interacting points will decay into a $\nu_\mu$ and $\bar\nu_e$, forming an intense and strongly collimated beam in the forward direction.
The $\nu_\mu$ could scatter in a downstream detector, placed $\sim 100$~m from the MuCol interaction point (IP), and produce the neutrinophilic scalar via the mono-neutrino process $\nu_\mu N \to \mu^+ \phi  X$ which could be detected through the presence of a wrong-sign muon in the final state.
The sensitivity for both a 10~kg and 10~ton mass detector as estimated in this work are shown the right panel of \cref{fig:schematic}. 
This clearly indicates the potential of a neutrino experiment at the muon collider to significantly exceed constraints obtained by past and planned searches and probe a variety of benchmarks scenarios in which the neutrinophilic mediator can explain the observed dark matter relic abundance. 
At the same time, precision measurements of the Higgs-to-invisible decay rates at the main detector of a muon collider would provide complementary sensitivity to the same new-physics scenario.

The paper is structured as follows. In \cref{sec:model}, we present the neutrinophilic mediator model and possible connections to dark matter as well as existing constraints and sensitivities of other proposed searches. In \ref{sec:neutrino} we discuss how the neutrinophilic mediator can be probed by measuring neutrino interactions at a dedicated forward neutrino detector at a muon collider and obtain the resulting sensitivity. An alternative approach of testing neutrinophilic mediators using invisible Higgs decays at the MuCol is discussed in \cref{sec:Higgs}. We conclude in \cref{sec:conclusion}.

\section{Neutrinophilic Mediator Model}
\label{sec:model}

In this work, we consider an additional neutrinophilic particle $\phi$ which couples neutrinos to one or more dark matter particles. Additionally, this mediator can lead to stronger-than-expected neutrino self-interactions. We begin with the assumption that $\phi$ carries lepton number $-2$ and its interactions with the SM arise from a dimension-six operator~\cite{Berryman:2018ogk},
\begin{align}
    \label{eq:Dim5Lagrangian}
    \mathcal{L} \supset \frac{1}{\Lambda_{\alpha\beta}^2} \left(L_\alpha H\right) \left(L_\beta H\right) \phi + \mathrm{h.c.},
\end{align}
where $\Lambda_{\alpha\beta}$ is the scale at which this EFT breaks down, and we implicitly allow for some flavor dependence. After electroweak symmetry breaking, this generates several interaction terms in the Lagrangian -- of particular interest here is the term
\begin{align}
    \mathcal{L} \supset \frac{1}{2} \lambda_{\alpha\beta} \nu_\alpha \nu_\beta \phi + \mathrm{h.c.},
\label{eq:Lagrangian}
\end{align}
where $\lambda_{\alpha\beta} = v^2/\Lambda_{\alpha\beta}^2$. Simultaneously,~\cref{eq:Dim5Lagrangian} generates an interaction term $\mathcal{L} \supset \left(\lambda_{\alpha\beta}/v\right) \nu_\alpha \nu_\beta \phi h$, which allows for Higgs boson decays into $\phi$ and a pair of neutrinos, assuming $m_\phi < m_h$. For the remainder of this work, we will assume for simplicity that $\lambda_{\mu\mu}$ is the only nonzero coupling -- further discussion of alternative flavor-structure can be found in Ref.~\cite{Berryman:2022hds}.

Assuming that $\phi$ also couples to dark matter, it may mediate dark matter/SM interactions and could set the dark matter relic abundance. We consider two scenarios. 
\begin{description}
\item[Thermal freeze-out Dark Matter] 
Since $\phi$ carries lepton-number $-2$, dark matter particles interacting to $\phi$ must be charged as well. In some cases, this guarantees the stability of the dark matter~\cite{Kelly:2019wow}. Two such examples, explored thoroughly in Ref.~\cite{Kelly:2019wow}, are that the dark matter is a Dirac fermion (DF) with lepton-number $+1$ or a complex scalar (CS) with lepton-number $+2/3$. In these two cases, the relevant interaction terms in the Lagrangian are
\begin{align}
    \label{eq:DiracFermionLagrangian}
    \mathcal{L}_{\rm DF} \supset \frac{1}{2}y \bar\chi^c \chi \phi + \mathrm{h.c.},\quad \mathcal{L}_{\rm CS} \supset \frac{1}{6}y \chi^3 \phi + \mathrm{h.c.}.
\end{align}
In both of these cases, with sufficiently strong neutrino-DM interactions, the dark matter will thermalize with the SM in the early universe. As the temperature of the plasma drops close to the DM mass, the DM-neutrino interactions will freeze out, and if the thermally-averaged cross section $\sigma v (\chi \chi \to \nu \nu)$ (DF) or $\sigma v(\chi \chi \to \chi^* \nu\nu)$ (CS) is of the correct size, the relic abundance today of $\chi$ could be responsible for the observed abundance of dark matter. For a given $m_\phi$ and $m_\chi$ this amounts to the product $y\lambda_{\alpha\beta}$ taking on a particular value. Ref.~\cite{Kelly:2019wow} found that, for ${\sim}$GeV-scale masses for $\chi$ and $\phi$, this amounts to $y\lambda_{\alpha\beta} \approx 10^{-2}$. Lines of constant $\Omega_\chi h^2$ satisfying the observed relic abundance of dark matter are shown for the DF (black, dot-dashed line) and CS (black, dotted line) cases in~\cref{fig:schematic}(right). Here, we assume $y=1$ and $m_\phi = 3m_\chi$ -- different assumptions, e.g., a different ratio of $m_\phi/m_\chi$, can cause these relic-density targets to shift slightly. For example, Refs.~\cite{Kelly:2019wow} demonstrated that if instead, $m_\phi = 10m_\chi$, the relic-density curves shift slightly upward towards larger $\lambda_{\mu\mu}$. For the analyses of interest in this work, the sensitivity depends solely on $\lambda_{\mu\mu}$ and $m_\phi$, independent of $m_\chi$ or $y$.

\item [Freeze-in Sterile-neutrino Dark Matter]
An alternative possibility related to DM is that the mediator $\phi$ induces relatively large neutrino self-interactions that, coupled with mixing between the SM neutrinos and a new sterile neutrino $\nu_s$, allows an abundance of $\nu_s$ to be populated over the thermal evolution of the universe. This is akin to the Dodelson-Widrow mechanism~\cite{Dodelson:1993je}, augmented by these stronger self-interactions and has been explored, for instance, in Refs.~\cite{deGouvea:2019wpf,Kelly:2020aks,Kelly:2020pcy}. In this regard, sterile neutrinos with ${\sim}$10~keV masses and relatively small mixing with the SM can be adequately populated as DM with relatively large $\lambda_{\alpha\beta}$ that induce the self-interactions. However, this form of DM is in tension with the abundances of Milky-Way satellite counts~\cite{An:2023mkf}. Nevertheless, we view this freeze-in scenario as an interesting target for searches for the neutrinophilic mediator $\phi$.

\end{description}

The interaction in~\cref{eq:Lagrangian} leads to neutrino-neutrino-$\phi$ vertices -- however, since $\phi$ carries lepton number, the character of the neutrino interacting with $\phi$ ``flips'' from neutrino to antineutrino, or vice versa. In the following sections, we explore how this vertex allows for nontrivial neutrino-nucleus scattering~\cite{Berryman:2018ogk,Kelly:2019wow} that can be searched for in a variety of environments -- specifically for this work, a muon-collider-sourced neutrino detector.

The neutrinophilic $\phi$ may be searched for in numerous ways when $\lambda_{\mu\mu} \neq 0$ -- we summarize several of the leading constraints as a function of $m_\phi$ here, presented as shaded grey regions in~\cref{fig:schematic} (right). For masses $m_\phi \lesssim$~MeV, $\phi$ can be thermalized with the SM and remain in contact throughout big-bang nucleosynthesis. If so, $\phi$ will act as (approximately) an additional degree of relativistic radiation throughout light-element formation, which is severely constrained for $m_\phi < 2$~MeV~\cite{Blinov:2019gcj}. On the other hand, neutrino self interactions can also leave imprints on the Cosmic Microwave Background (CMB). The gray shaded region in the parameter space is excluded by the current CMB data ~\cite{Barenboim:2008ds}. The parameter space is also constrained by rare decays of $Z\rightarrow\nu_\mu\nu_\mu\phi$~\cite{Brdar:2020nbj} and mesons $M^{\pm}\rightarrow \mu\nu_\mu\phi$~\cite{Berryman:2018ogk, Dev:2024twk} by the PIENU~\cite{PIENU:2021clt} and NA62~\cite{NA62:2021bji} for rare $\pi^\pm$ and $K^\pm$ decays, respectively. 

Detecting high-energy astrophysical neutrinos with IceCube has provided a unique avenue to explore neutrino self-interactions, despite current constraints being weaker than laboratory probes. Future observatories like IC-Gen2 promise significant improvements in sensitivity~\cite{Esteban:2021tub}. The DUNE facility which will have an intense beam of neutrinos (but largely uncontaminated by antineutrinos) with energies of a few GeV and will be able to study neutrino BSM signature as discussed in this paper looking for wrong sign muon with the help of liquid argon near detector~\cite{Kelly:2019wow, Berryman:2018ogk}. On the other hand, in the far forward region of LHC, at the FLArE detector at the Forward Physics Facility (FPF), the same BSM signature will be probed~\cite{Kelly:2021mcd,Bai:2024kmt}. However, due to the presence of both neutrinos and antineutrinos in the beam, FLArE will mainly look for missing energy signatures. Another interesting channel to look for the scalar mediator is $h\rightarrow\nu\nu\phi$ and $h\rightarrow\bar\nu\bar\nu\phi$ decays. Both LHC at High Luminosity era~\cite{deGouvea:2019qaz} and MuCol as discussed in the upcoming section will be able to probe this decay channel. In our analysis, we focus on a MuCol colliding opposite-sign muons -- a same-sign muon collider~\cite{Hamada:2022mua} has very strong sensitivity in the same parameter space we explore by searching for signals of apparent lepton-number violation and missing energy, as emphasized in Ref.~\cite{deLima:2024ohf}.

\section{Neutrino Scattering Signals}
\label{sec:neutrino}

In this section, we focus on the neutrino-induced signature related to the BSM scalar introduced in~\cref{eq:Lagrangian}. We first briefly discuss the neutrino detector concept used in our analysis. We then present the new physics signal and backgrounds and provide a detailed analysis of how to discriminate between them.

\subsection*{Neutrino Detection at the Muon Collider}

Any circular muon collider naturally consists of a muon storage ring that would deliver the beam to the collision points. Due to their short lifetime, the stored muons will decay and produce a large number of neutrinos. This makes a muon collider a powerful neutrino source.

In this study, we will use the neutrino fluxes obtained for a muon collider with a center-of-mass energy of 3~TeV as presented in Ref.~\cite{InternationalMuonCollider:2024jyv}. This design envisions a 4.5~km long storage ring into which about $10^{13}$ muons will be injected each second. While these muons can decay anywhere in the ring, the largest flux of neutrinos originates from decays in the straight sections around the IPs. We conservatively assume that the straight section have a length of 10~m, resulting in about $10^{10}$ neutrinos per second. This exceeds the flux predictions for ongoing and future forward neutrino detectors at hadronic colliders by several orders of magnitude~\cite{FASER:2019dxq, SNDLHC:2022ihg, Feng:2022inv, MammenAbraham:2024gun}.

The neutrinos produced in muon decay will have a transverse momentum similar to the muon mass and typical energies of 100~GeV or more. The neutrino beam is then highly collimated with an angular spread of $m_\mu/E_\nu \lesssim 1$~mrad. Therefore, even a relatively small cylindrical detector with radius $R = 10~\cm$ placed 100~m downstream from IP and centered on the muon beam collision axis is sufficient to encompass the entire beam. Detectors placed further away with correspondingly larger transverse size would lead to the same sensitivity.  

Crucially, the MuCol$\nu$ detector placed on one side of the IP at the collider will capture essentially pure samples of muon neutrinos $\nu_\mu$ and electron antineutrinos $\bar{\nu}_e$, while the corresponding antiparticles, $\bar{\nu}_\mu$ and $\nu_e$, will be produced in the opposite direction from the IP. This directional separation plays an important role in mitigating backgrounds, as discussed below; see also Ref.~\cite{Batell:2024cdl} for a similar discussion in the context of a muon fixed-target experiment at the MuCol. Notably, forward neutrinos at the MuCol have a narrower energy spread than those at hadronic colliders, and their spectrum can be predicted with greater accuracy (given that they come from the well-known three-body muon decay), reducing the corresponding systematic uncertainties in neutrino measurements. We assume that these uncertainties can be further suppressed by measuring the dominant charge current (CC) neutrino interaction rates, rendering them negligible in our analysis.

A preliminary neutrino detector concept for muon storage rings and colliders, proposed in Ref.~\cite{King:1997dx}, consists of a $1~\textrm{m}$ long cylindrical detector with a $10~\textrm{cm}$ radius. It comprises $750$ silicon CCD tracking planes followed by a magnetized spectrometer and calorimeter. This design aims to ensure strong tracking and reconstruction capabilities, as well as $c$/$b$-quark tagging. In this concept, the silicon target mass used for neutrino vertex detection is only $10~\textrm{kg}$. However, interleaving the active target with denser materials could significantly increase this mass and, consequently, the expected neutrino interaction rate.

We present our results for two representative MuCol$\nu$ detector masses: $10~\textrm{kg}$ and $10~\textrm{ton}$. For each, we study the expected BSM signal and backgrounds per year of MuCol operation. The larger detector mass is equivalent to, for example, a $1~\textrm{ton}$ detector operating for $10$ years, accumulating the same total number of interactions. This $1~\textrm{ton}$ target mass is similar to that of the currently operating forward neutrino detectors at the LHC~\cite{FASER:2023zcr, SNDLHC:2023pun, FASER:2024hoe, Ariga:2025qup}. For simplicity, we assume iron target nuclei in our analysis. However, the results are largely driven by the detector mass and only mildly dependent on the specific nucleus. 

\subsection*{Signal and Background}

Our key signature is the process $\nu_\mu N \to \mu^+ \phi X$ in the MuCol$\nu$ detector, as illustrated in~\cref{fig:schematic}; see also Ref.~\cite{Liu:2024ywd} for projected bounds based on a missing-energy signature in the MuCol main detector. This process, and many of the key observables with which it can be separated from neutrino-scattering backgrounds, has been explored in detail in Refs.~\cite{Berryman:2018ogk,Kelly:2019wow,Kelly:2021mcd}. In this section, we emphasize the ways in which the MuCol neutrino source and the MuCol$\nu$ detector excel in a search of this type. For instance -- the scalar carries lepton number and escapes detection, so the interaction mimics a lepton-number-violating process. Because of the pure $\nu_\mu$ beam produced by the MuCol (without contamination from $\bar\nu_\mu$), as well as the magnetization of the MuCol$\nu$ spectrometer, we expect exquisite capabilities using this approach.

We calculate the relevant cross sections using \texttt{MadGraph5\_aMC@NLO v2.9}~\cite{Alwall:2014hca} and employ the \texttt{nCTEQ15} nuclear parton distribution function (PDF) set for the iron target nucleus~\cite{Kovarik:2015cma}. \texttt{Pythia8}~\cite{Sjostrand:2014zea} is used to further model the parton shower and hadronization for the signal events. We have validated the numerical results against a semi-analytical estimates based on the partonic cross sections provided in Refs.~\cite{Kelly:2021mcd}. For small scalar masses and transverse momenta $m_\phi\ll, p_{T_\phi} \sqrt{s}$, the scalar production cross section in neutrino-nucleon scattering approximately scales as
\begin{equation}
\sigma(\nu_\mu n \to \mu^+ \phi X ) \simeq 10^{-37}~\textrm{cm}^2 \times \lambda^2 \times \frac{E_\nu}{\textrm{TeV}}\, ,
\label{eq:sigmaapprox}
\end{equation}
where $\lambda$ is the scalar coupling to neutrinos and $E_\nu$ is the incident neutrino energy.  

\texttt{Pythia8} is also employed to model neutrino-induced backgrounds. These backgrounds include charged current and neutral current (NC) scatterings of $\nu_\mu$ and $\bar{\nu}_e$ that can occasionally lead to positive muons in the final state, among other visible products. This is dominantly due to neutrino-induced charm meson production with subsequent decays into muons. Approximately $10\%$ of all CC $\nu_\mu$ interactions at multi-100~GeV energies have a charm quark in the final state, $\nu_\mu s \to \mu^- c$, and about $10\%$ of charm hadrons decay semileptonically to muons, $D^+, D^0 \to \mu^+ + X$~\cite{ParticleDataGroup:2024cfk}. Hence, roughly $1\%$ of all CC $\nu_\mu$ events are expected to produce a positively charged muon. In contrast, the suppression is larger for NC events and CC $\bar{\nu}_e$ scatterings, as the lower $c$ quark PDF suppresses the production rate of charm quarks in these processes. An additional source of positively charged muons are rare light meson decays, such as $\eta \to \mu^+\mu^-\gamma$ or $\omega \to \mu^+ \mu^-$. However, these particle do not necessarily carry a large momentum fraction and the muons therefore tend to be less energetic. Crucially, all these backgrounds initially dominate over the expected BSM signal by several orders of magnitude. Below, we discuss in detail a possible background mitigation strategy that allows for isolating new physics events.

In addition to backgrounds with positive muons, neutrino interactions can mimic signal events through other processes. For instance, charged pion $\pi^+$ mis-reconstruction could lead to a false muon signal. We assume that the detector's muon identification and charged pion discrimination capabilities are sufficient to suppress such backgrounds. Furthermore, charged pions or kaons can decay into muons within the detector before being identified. To reject these backgrounds, sufficient angular resolution is necessary to identify in-flight pion decays by observing the momentum kink between the incident pion and outgoing muon tracks. While such detector capabilities could, in principle, mitigate these backgrounds, they remain sensitive to a detailed detector design. Hence, in the following, we will include these \textsl{displaced backgrounds} from $\mu^+$ produced from charged pion and kaon decays in flight, on top of the previously mentioned \textsl{prompt backgrounds} from $\mu^+$ production at the neutrino interaction vertex. When modeling the former backgrounds, we require charged hadrons to decay inside the spectrometer, which is assumed to be 10 m long, before they reach the calorimeter and the muon ID system of the detector.

Neutrino event rates presented below are subject to uncertainties in modeling $\nu$ interactions, particularly the relevant cross section. These uncertainties, however, typically do not exceed a few percent for TeV-scale energies~\cite{Candido:2023utz}. They are also expected to be further reduced by the time the MuCol$\nu$ detector operates, thanks to the data gathered in the neutrino physics program at the LHC~\cite{Cruz-Martinez:2023sdv} as well as measurements at the planned electron-ion collider (EIC)~\cite{Accardi:2012qut}. Therefore, we assume these uncertainties will not significantly affect the considered BSM search and neglect their impact in the following analysis. Finally, we include the effects of finite energy measurement resolution by smearing the momenta of outgoing particles produced in neutrino interactions. We assume Gaussian smearing with a $10\%$ relative uncertainty on the measured energy for all particles.

\subsection*{Analysis}

\begin{table*}[thb]
\setlength{\tabcolsep}{6pt}
\begin{tabular}{l | c c c c c | c c  }
\hline
\hline
& \multicolumn{5}{c|}{Background Rates} & \multicolumn{2}{c}{Signal Efficiency for $m_\phi$} \\
Cut & CC prompt & CC displaced & NC prompt & NC displaced & All & ~~1~GeV~~ & \!\!\!20~GeV\!\!\! \\
\hline
All Events  & \multicolumn{2}{c|}{$1.89 \cdot 10^{7}$}  & \multicolumn{2}{c|}{$6.24 \cdot 10^{6}$}  & $2.52 \cdot 10^{7}$  & $1.00 $  & $1.00 $\\
\cline{2-6}
$E_{\mu^+}>100$~GeV  & $3.23 \cdot 10^{4}$  & $4.59 \cdot 10^{3}$  & $2.45 \cdot 10^{3}$  & $1.29 \cdot 10^{3}$  & $4.06 \cdot 10^{4}$  & 0.783  & 0.661\\
$E_{\mu^-}<30$~GeV  & $2.43 \cdot 10^{3}$  & $9.11 \cdot 10^{2}$  & $2.30 \cdot 10^{3}$  & $1.29 \cdot 10^{3}$  & $6.93 \cdot 10^{3}$  & 0.78  & 0.661\\
$E_{\mu^+} > 0.5 E_{visible}$  & $2.40 \cdot 10^{2}$  & 33.77  & $2.09 \cdot 10^{2}$  & $1.79 \cdot 10^{2}$  & $6.61 \cdot 10^{2}$  & 0.576  & 0.412\\
charm veto  & 48.47  & 30.85  & 43.87  & $1.78 \cdot 10^{2}$  & $3.01 \cdot 10^{2}$  & 0.555  & 0.411\\
$p_{T,\mu^+}$ vs. $\Delta \phi$  & 1.68  & 0.767  & 2.86  & 4.56  & 9.88  & 0.386  & 0.233\\
\hline
BDT  & 0.155  & 0.646  & 0.081  & 1.35  & 2.23  & 0.416  & 0.214\\
\hline
\hline
\end{tabular}
\caption{Event yields for background processes and the signal efficiency after consecutive cuts obtained for a $10~\textrm{kg}$ MuCol$\nu$ detector placed in the forward kinematic region of the MuCol and one year of data collection. Results are shown for prompt and displaced background contributions, and for two masses of the neutrinophilic scalar, $m_\phi = 1$ and $20~\textrm{GeV}$. The CC and NC prompt and displaced categories include events from both $\nu_\mu$ and $\bar{\nu}_e$ interactions.}
\label{tab:cutflow}
\end{table*}

As discussed earlier, neutrino-induced backgrounds dominate over the BSM signal and must be rejected in the analysis. To illustrate how background can be suppressed, we first perform a simple-cut-and count analysis using several key observables. To further improve the rejection and optimize the analysis performance, we also perform a multivariate analysis using a Boosted Decision Tree (BDT). The number of background events remaining after successive cuts is summarized in \cref{tab:cutflow}. These results correspond to a $10~\textrm{kg}$ MuCol$\nu$ detector operating for a year and are provided separately for CC and NC interactions of incident $\nu_\mu$ and $\bar{\nu}_e$. Both prompt and displaced backgrounds are considered. The table also shows the signal efficiency (fraction of surviving events) for two representative masses of the BSM scalar, $m_\phi = 1$ or $20~\textrm{GeV}$. 

As a first step, we identify candidate signal events by requiring the presence of a positively charged muon and the absence of a negatively charged muon:
\begin{description}
\item[Positive muon] We require at least one positive muon in the final state of the neutrino interaction. To ensure its proper identification and a reliable energy measurement, we require a sufficiently high muon energy, specifically $E_{\mu^+} > 100 ~\textrm{GeV}$. This allows the positive muon to escape the target material and traverse the magnetized spectrometer and the muon system, i.e., a dedicated detector component designed to identify muons and precisely measure their momenta. The magnet can deflect low-energy charged particles before they reach the muon system. This energy threshold is chosen based on a similar analysis for the FASER spectrometer at the LHC, which resembles the assumed MuCol$\nu$ detector setup~\cite{FASER:2018bac}.

Assuming a spectrometer employing a $10~\textrm{m}$ magnet with a field strength of $1~\textrm{T}$, a $1.5~\textrm{TeV}$ muon will bend by approximately $2~\textrm{cm}$ This estimate suggests that even a shorter spectrometer or weaker magnetic field could be sufficient for measuring the muon bending with modern detectors, which have resolutions on the order of $100~\mathrm{\mu m}$. To ensure the muon traverses the entire spectrometer and reaches the muon system, an energy threshold of $E_{\mu^+} \gtrsim 30$ GeV should be imposed. This threshold guarantees that the bending remains small enough for typical detector dimensions on the order of meters. In particular, we employ a $100~\textrm{GeV}$ threshold to maximize background and signal separation. For simplicity, we assume perfect muon sign identification within this setup and above the energy threshold. It is important to emphasize that muon sign identification is crucial for the analysis presented below. Future detector designs should guarantee excellent $\mu^+$ and $\mu^-$ separation to enable this search.

Notably, imposing a minimum energy threshold on $\mu^+$ has a limited impact on the BSM signal events, resulting in about $20-40\%$ suppression, depending on $m_\phi$. However, this threshold significantly reduces the background event rate. As discussed earlier, about $1\%$ of CC $\nu_\mu$ interactions produce a $\mu^+$ in the final state. Indeed, by requiring the presence of the positive muon, we observe a suppression of the corresponding prompt background event rate by more than two orders of magnitude. In this case, the positive muons often come from decays of leading charm hadrons, i.e., those with the largest momentum. Consequently, the energy threshold has a less pronounced impact on suppressing these events. For the other background types, the fraction of scattering events containing a charm quark are significantly lower and positive muon, originating from sub-leading charm hadrons or decays of lighter mesons produced in hadronic showers, are typically softer. In these cases, the requirement to have a muon with $E_{\mu^+}>100~\textrm{GeV}$ suppresses background event rates by more than three orders of magnitude.

\item[Negative muon] The leading background at this stage originates from CC $\nu_\mu$ interactions. Unlike the signal, these are expected to contain an energetic negatively charged muon. We therefore veto events containing negative muons with energies above $30~\textrm{GeV}$. This threshold is, again, chosen based on the assumed detector's ability to reliably identify muons. While soft negative muons ($E_{\mu^-}<30~\textrm{GeV}$) may be present in both signal and background events, their identification is less reliable, and thus no further cuts are applied on them.

While this requirement reduces the CC $\nu_\mu$ background by more than an order of magnitude, it has a negligible impact ($<1\%$) on the signal event rate. This minimal impact holds true for both $m_\phi = 1~\textrm{GeV}$ and $m_\phi = 20~\textrm{GeV}$, as the $\phi$ scalar, especially at higher masses, is expected to carry a significant portion of the incident neutrino energy, leaving less energy available for the visible final-state particles. 
\end{description}

These baseline identification conditions significantly reduce the background event rate (by almost four orders of magnitude) while maintaining a high signal detection efficiency (greater than $60\%$), as shown in \cref{tab:cutflow}. To further suppress backgrounds, we introduce additional cuts on the identified candidate signal events:

\begin{description}
\item[$\mathbf{E_{\mu^+}/E_{\textrm{vis}} > 0.5}$] In the BSM signal events, the positive muon originates from the leptonic vertex in neutrino-nucleus scattering, distinct from $\mu^+$ production in the hadron decays characteristic for background events. Consequently, one expects that in signal events, $\mu^+$ typically carries a more significant fraction of the total visible energy, $E_{\textrm{vis}}$.

This is illustrated in the top left panel of \cref{fig:1D}, which shows the distribution of interacting neutrinos in bins of the ratio $E_{\mu^+}/E_{\textrm{vis}}$. Prompt and displaced background events from CC and NC neutrino scatterings are shown in color. These events have satisfied the baseline conditions on the presence and absence of positive and negative muons. For comparison, the corresponding distribution for the BSM scalar with $m_\phi = 1~\textrm{GeV}$ and $\lambda = 0.3$ is also shown.

\begin{figure*}[thb]
\centering
\includegraphics[width=0.45\textwidth]{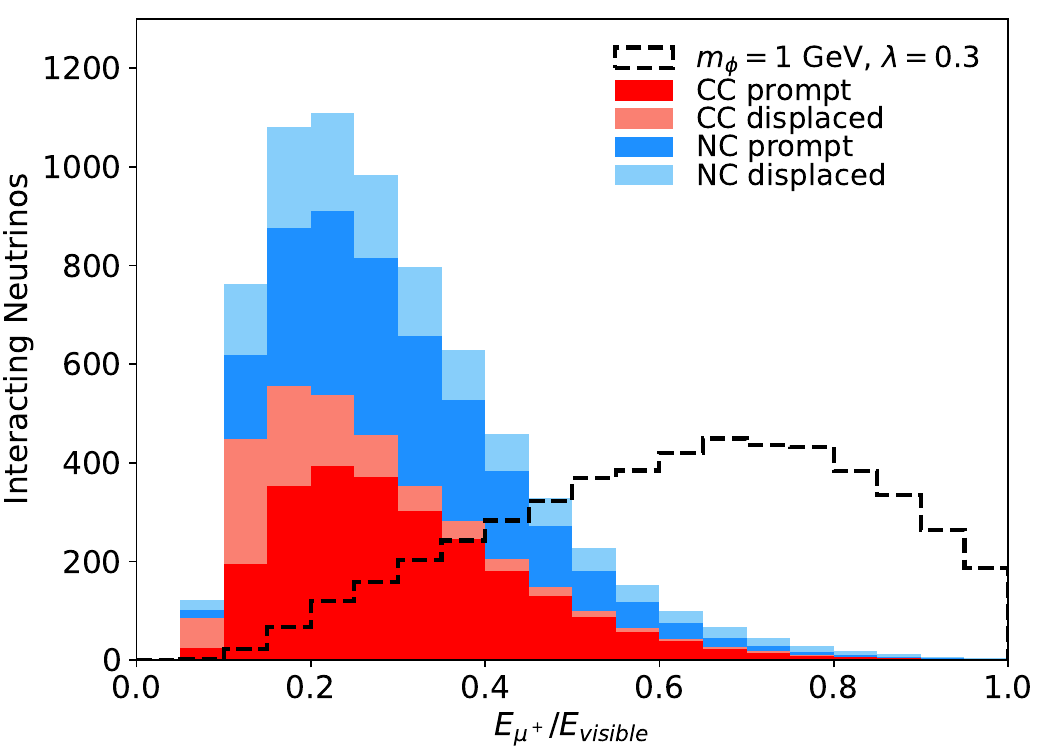}
\includegraphics[width=0.45\textwidth]{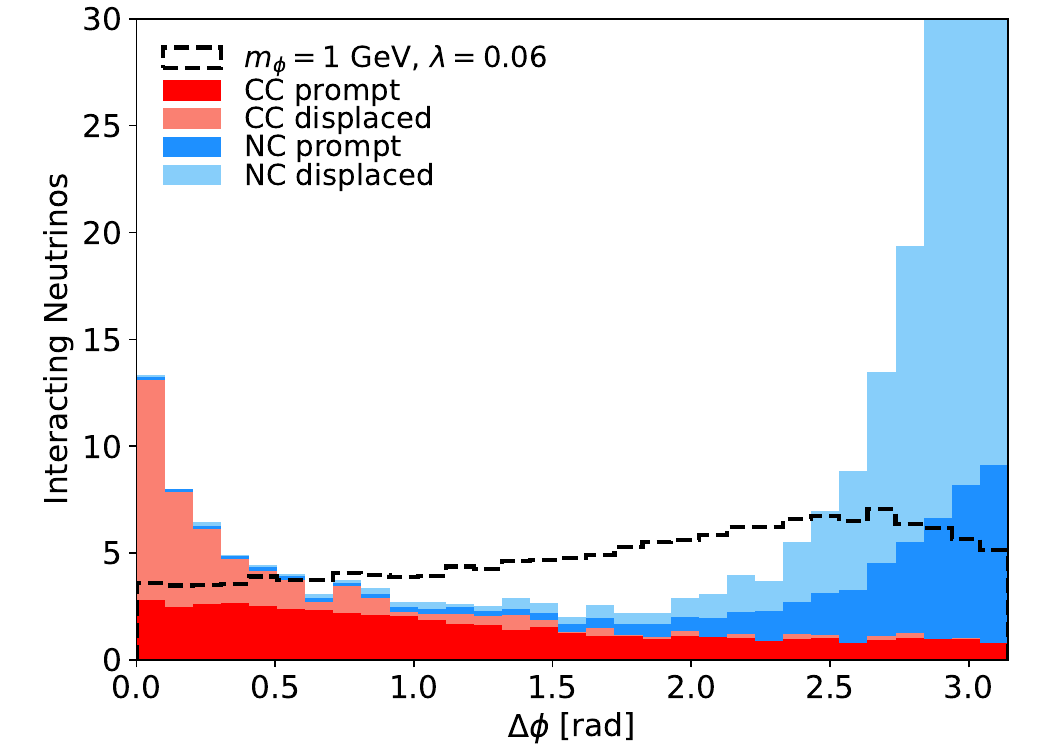}
\includegraphics[width=0.45\textwidth]{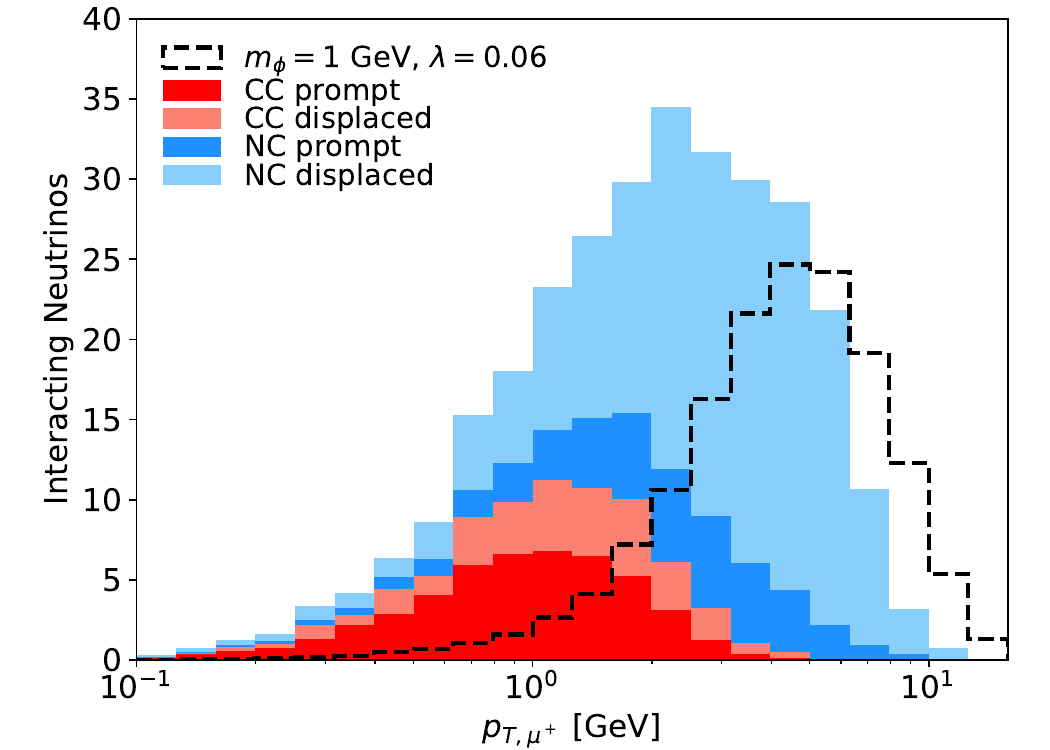}
\includegraphics[width=0.45\textwidth]{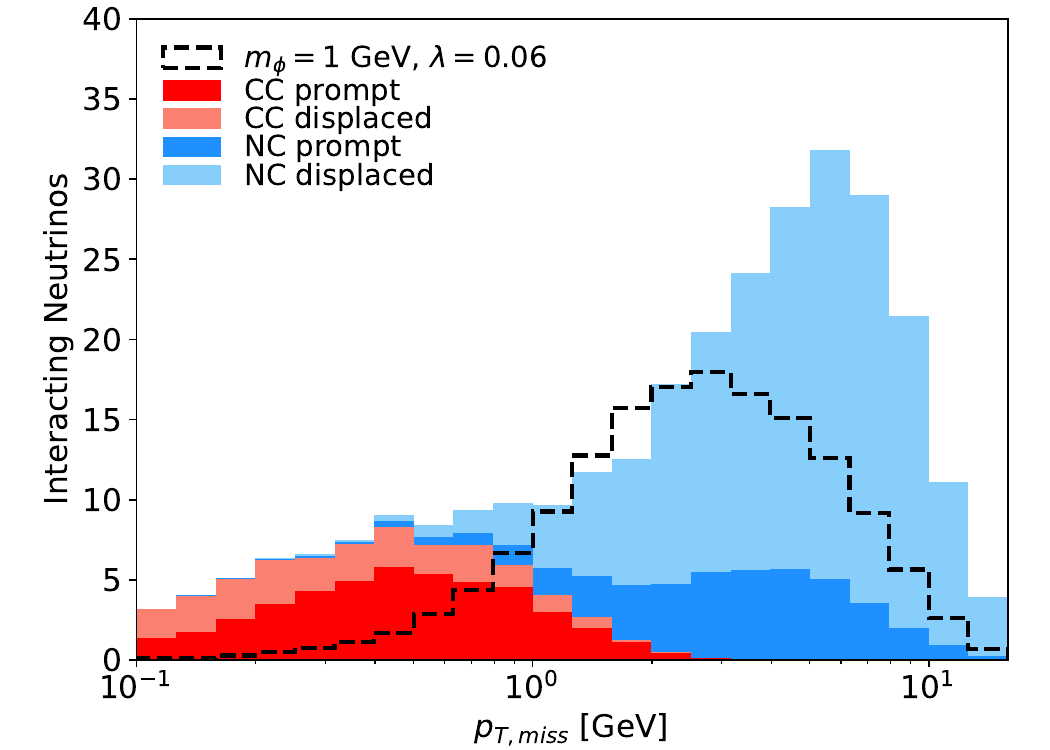}
\caption{Distributions of selected kinematic variables used in the analysis of neutrino-nucleus interactions within a proposed $10~\textrm{kg}$ MuCol$\nu$ detector at a $3~\textrm{TeV}$ muon collider. Stacked histograms show the predicted number of events per year. The distributions are shown for: the ratio between the positive muon and total visible energies, $E_{\mu^+}/E_{\textrm{vis}}$ (\textbf{top left}); the angle between the $\mu^+$ and missing transverse momenta, $\Delta\phi$ (\textbf{top right}); the transverse momentum of the positive muon, $p_{T,\mu^+}$ (\textbf{bottom left}); and the missing transverse momentum, $p_{T,\textrm{miss}}$ (\textbf{bottom right}). Charged-current (CC) and neutral-current (NC) neutrino-induced background events are shown in blue and red, respectively and stacked in histograms. Dark and light shading within the histograms differentiates between muon neutrinos $\nu_\mu$ and electron antineutrinos $\bar{\nu}_e$. Sample distributions for BSM signal events, indicated by black dashed lines, assume a scalar mass $m_\phi = 1~\textrm{GeV}$ and a coupling strength $\lambda = 0.3$ (top left) or $0.06$ (other panels).}
\label{fig:1D}
\end{figure*}

When defining the total visible energy, $E_{\textrm{vis}}$, we assume that the energy of all the charged particles, electromagnetic showers induced by neutral pion or $\eta$ decays, and neutral hadrons can be measured in the spectrometer and calorimeter placed downstream to the target material. The charged particles and electromagnetic showers correspond to a majority of energy deposited in the detector. However, measuring a subdominant fraction of neutral hadrons is also important to achieve the $10\%$ resolution on $E_{\textrm{vis}}$. Crucially, this necessitates the employment of a hadronic calorimeter with good containment of hadronic showers at energies of order tens of GeV. Based on the detector concept from Ref.~\cite{King:1997dx}, we assume that a hadronic calorimeter with these capabilities will be installed. The asssumed feasibility of such measurement corresponds to the expected progress in calorimetry, cf. review prepared for the proposed Future Circular Collider (FCC)~\cite{Aleksa:2021ztd}.

The ratio $E_{\mu^+}/E_{\textrm{vis}}$ provides strong discrimination between signal and background events. In particular, requiring $E_{\mu^+}/E_{\textrm{vis}} > 0.5$ reduces the background event rate by more than one order of magnitude. We expect a few hundred background events to survive this cut and mimic the signal.

The impact of the cut on signal efficiency depends on the BSM scalar mass, ranging from approximately $60\%$ to $40\%$ for the benchmark masses of $1$ and $20~\textrm{GeV}$, respectively. Notably, similar background rejection and signal detection efficiencies would not be achieved by using the positive muon energy $E_{\mu^+}$ or the visible energy $E_{\textrm{vis}}$ alone. Both quantities decrease with increasing BSM scalar mass, as $\phi$ carries away a progressively larger fraction of the incident neutrino energy. Consequently, the efficiency of such cuts depends more strongly on $m_\phi$. This dependence is partially mitigated by using the ratio of the two quantities in the analysis.

\item[Charm tagging] As discussed above, energetic positive muons in background events are typically associated with a parent hadron carrying even higher energy. Sufficiently long-lived hadrons, such as charm mesons, can be identified in the detector based on their displaced decay. Therefore, charm tagging provides additional discrimination against prompt backgrounds.

The proposed MuCol$\nu$ detector incorporates multiple high-resolution tracking planes capable of identifying these displaced vertices. In our analysis, we apply a charm tagging requirement with an assumed efficiency of $80\%$ for hadrons decaying to $\mu^+$. Events induced by $B$-meson and $\tau$-lepton decays are similarly rejected. The mistagging rate is assumed to be negligible for the BSM signal rate estimation. This selection reduces prompt backgrounds by approximately a factor of two, while it has a minimal impact on displaced backgrounds. Therefore, as we will demonstrate later, charm tagging does not play a dominant role in our multivariate analysis, which is significantly enhanced by the use of the BDT.


\item[Transverse observables] Massive BSM neutrinophilic scalars can naturally be produced with substantial transverse momentum, $p_T\sim m_\phi$, which can be measured even in interactions of high-energy neutrinos. The presence of this missing transverse momentum alters the kinematics of $\nu$ scattering and can be used to discriminate against backgrounds.

In particular, the top right panel of \cref{fig:1D} shows the distribution of the angle $\Delta\phi$ between the transverse momentum of the final-state positive muon, $p_{T,\mu^+}$, and the missing transverse momentum, $p_{T,\textrm{miss}}$. In the background NC events, $p_{T,\textrm{miss}}$  is dominated by the outgoing neutrino, which recoils against the hadronic system containing the final-state $\mu^+$. Consequently, this angle exhibits a strong peak near $\Delta\phi\sim \pi$, indicating a preference for back-to-back topology in the transverse plane. In contrast, BSM signal events have a much broader $\Delta\phi$ distribution, given the presence of both energy associated with the undetected $\phi$ and $\mu^+$ emerging from the leptonic vertex in the neutrino interaction. Signal events can then be effectively discriminated from the NC backgrounds by rejecting events with back-to-back topology. Additionally, a softer peak near 
$\Delta\phi\sim 0$ arises primarily from charged current (CC) events, where the missing transverse momentum is smaller, leading to a more collinear topology, which can also be rejected in the analysis.


\Cref{fig:1D} also shows the individual $p_{T,\mu^+}$ and $p_{T,\textrm{miss}}$ distributions in the bottom panels. These distributions illustrate that both variables could be used to further disentangle the BSM signal and CC background events, especially for prompt backgrounds (blue histograms). The distribution of the muon transverse momentum appears to be a more effective discriminant in this analysis. In addition, it also exhibits a weaker dependence on $m_\phi$.

We examine the correlations between $\Delta \phi$ and $p_{T,\mu^+}$, shown in \cref{fig:2D} to improve signal detection and background rejection efficiency further. Since background $\nu$-induced events typically have lower  $p_{T,\mu^+}$ than the BSM signal, and NC events with high $p_{T,\mu^+}$ prefer $\Delta\phi\sim \pi$, we isolate signal events by focusing on the following kinematic region: $p_{T,\mu^+} > 5~\textrm{GeV} \times e^{-\Delta \phi}$, $p_{T,\mu^+} > 2~\textrm{GeV}$ and $p_{T,\mu^+} > 85~\textrm{MeV} \times e^{1.5\Delta \phi}$. This reduces the background event rate by an additional order of magnitude to the level of $\mathcal{O}(10)$ events in the $10~\textrm{kg}$ detector per year. 
\end{description}

\begin{figure*}[thb]
\centering
\includegraphics[width=1\textwidth]{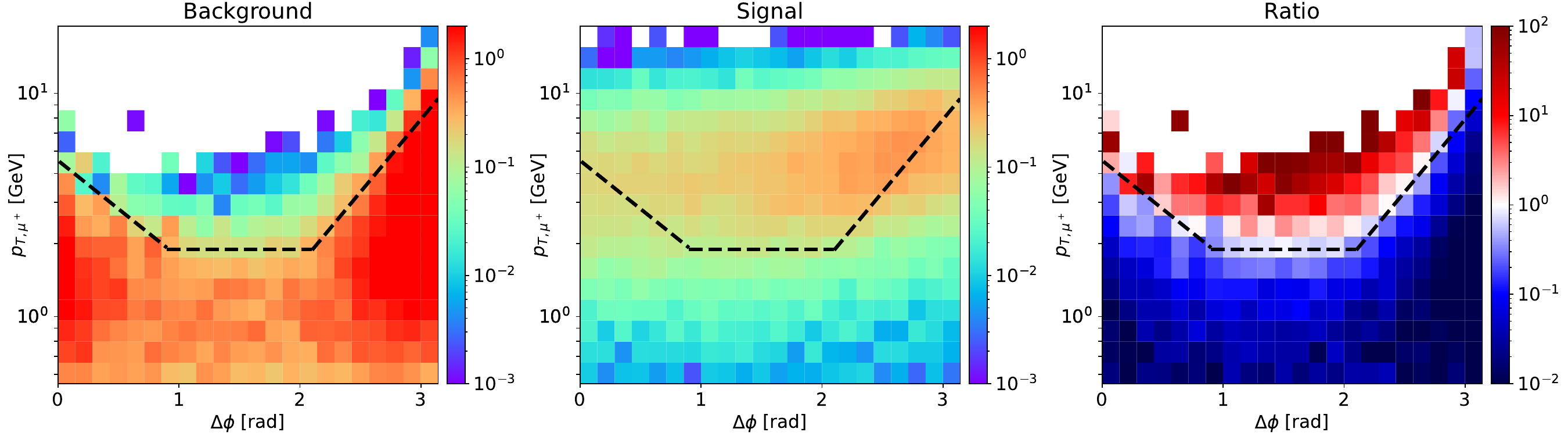}
\caption{Two-dimensional distributions of transverse observables used in the analysis, depicting the correlation between the angle between the $\mu^+$ and missing transverse momenta, $\Delta\phi$, and the positive muon transverse momentum, $p_{T,\mu^+}$.  The figure compares background events (\textbf{left}) to BSM signal events (\textbf{center}) obtained assuming $m_{\phi} = 1~\textrm{GeV}$ and $\lambda = 0.03$. The ratio of signal to background events is also presented (\textbf{right}).}
\label{fig:2D}
\end{figure*}

The overall background suppression of the presented analysis corresponds to more than six orders of magnitude, while the signal detection efficiency can be kept high, of about $20-40\%$, depending on the $\phi$ mass. While this analysis demonstrates how signal and SM background can be separated using simple physics-motivated cuts, a optimized multivariate analysis that considers additional correlations between the observables can further improve signal acceptance and background rejection.

\begin{figure*}[thb]
\centering
\includegraphics[width=0.45\textwidth]{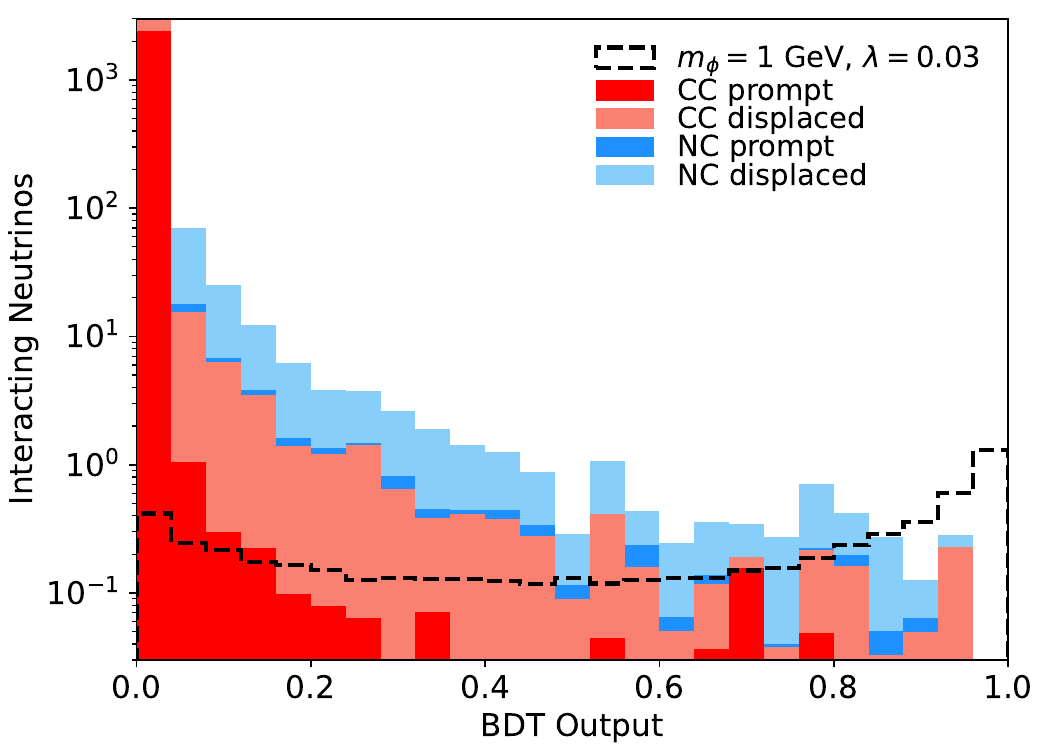}
\includegraphics[width=0.45\textwidth]{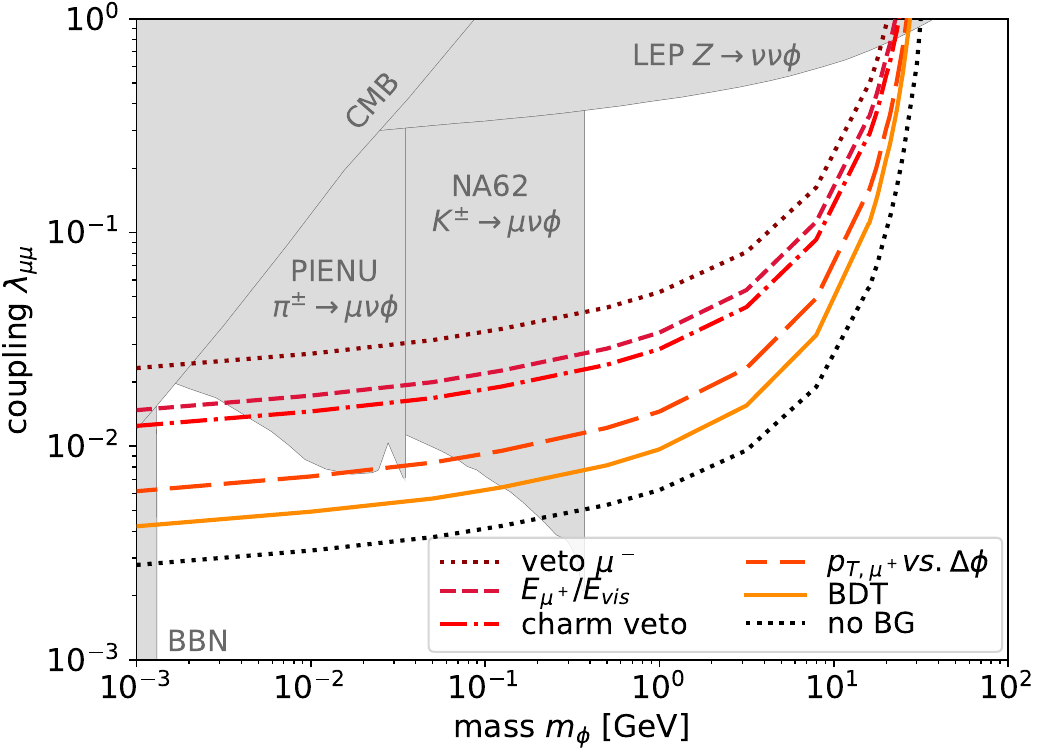}
\caption{\textbf{Left: BDT output variable distribution.} Similar to \cref{fig:1D}, but the distributions of the BDT score are shown for background and signal events. The latter correspond to the neutrinophilic scalar with $m_\phi = 1~\textrm{GeV}$ and $\lambda = 0.03$. \textbf{Right: Projected exclusion bounds for various cuts.} Projected exclusion bounds obtained for the $10~\textrm{kg}$ MuCol$\nu$ detector at the Muon Collider operating for one year. From top to bottom, the lines represent the impact of adding consecutive cuts on kinematic variables discussed in the text, as indicated in the plot. The best-expected constraints are obtained with the BDT analysis and shown with the orange solid line. The black dotted line represents an ultimate limit obtained, assuming that backgrounds can be suppressed to negligible levels without affecting the number of signal events.}
\label{fig:BDTandcuts}
\end{figure*}

\begin{description}
\item[BDT] To this end, we employ the BDT implemented in the \textsc{scikit-learn} package~\cite{scikit-learn}. Candidate signal events are preselected to have at least one $\mu^+$ with $E_{\mu^+} \gtrsim 100~\textrm{GeV}$ but no $\mu^-$ with energy $E_{\mu^-}>30~\textrm{GeV}$, as discussed above. For each event, the following observables are considered: \textsl{i)} the muon energy $E_{\mu^+}$; \textsl{ii)} the visible energy $E_\text{vis}$; \textsl{iii)} the energy deposited in electromagnetic showers $E_\text{EM}$; \textsl{iv)} the positive muon transverse momentum $p_{T,\mu^+}$; \textsl{v)} the missing transverse momentum $p_{T,\text{mis}}$; \textsl{vi)} the scalar sum of all visible final state particle transverse momenta $H_T$; \textsl{vii)} the angle between the positive muon and missing transverse momenta $\Delta \phi$; and \textsl{viii)} the condition whether or not the event was charm tagged (assuming a $80\%$ charm tagging efficiency). The BDT was configured with 200 estimators, a maximal depth of 3, and a learning rate of 0.03. These hyperparameters were determined through a coarse scan, but no further optimization was performed. The event set was split equally into training and testing samples, with similar performance observed for both.

As shown in the left panel of \cref{fig:BDTandcuts}, the BDT output variable provides good separation between signal and background. Applying a cut at $>0.7$ on this variable reduces the background event rate to only a few events, cf. \cref{tab:cutflow}, while maintaining a signal efficiency between $40\%$ (low mass of $\phi$) and $20\%$ (high mass). In the case of the $10~\textrm{ton}$ detector, due to larger event rates, a stronger cut of $>0.90$ was chosen in order to guarantee better discrimination power.

Interestingly, a feature importance analysis reveals that the separation is strongly driven by the transverse observables $p_{T,\mu^+}$ and $\Delta \phi$ as well as the muon energy $E_{\mu^+}$. Charm tagging plays a less important role, albeit not negligible, and it would not be critical for the successful search. This is an important consideration for potentially planning a larger $10$-ton detector, which may have limited charm tagging capabilities. As the BDT study demonstrates, this limitation should not hinder the discovery potential of such a detector in the search for neutrinophilic scalars.
\end{description}

\subsection*{Sensitivity \& Discussion}

The right panel of \cref{fig:BDTandcuts} illustrates the projected exclusion bounds on the neutrinophilic scalar, obtained with a $10~\textrm{kg}$ MuCol$\nu$ detector and one year of MuCol operation at a center-of-mass energy of $3~\textrm{TeV}$. To derive these bounds, we require a signal-to-background ratio satisfying $S/\sqrt{B} > 2$ and a minimum of three signal events, $S>3$.

The plot demonstrates the impact of consecutive cuts introduced in the analysis. Although the number of neutrino-induced background events is initially large, it can be significantly suppressed by imposing simple signal selection conditions on the muons ($\mu^+$ and $\mu^-$), as previously discussed. This alone allows us to probe new regions in the parameter space of the neutrinophilic scalar model, as shown by the red dotted line. Adding a further cut on the visible energy fraction carried by the positive muon, $E_{\mu^+}/E_{\textrm{vis}} > 0.5$, yields the improved bound indicated by the red short-dashed line. Rejecting charm-tagged events further enhances the sensitivity, leading to the bound depicted by the red dash-dotted line. Finally, incorporating constraints on transverse observables, cf. \cref{fig:2D}, produces the strongest bound from our cut-and-count analysis, represented by the red long-dashed line.

Employing the BDT yields even stronger bounds, as shown by the orange solid line. These constraints are only about a factor of two less stringent in $\lambda$ compared to the ultimate sensitivity of the $10~\textrm{kg}$ detector, illustrated by the black dotted line, which assumes perfect background rejection and signal detection efficiency.

\Cref{fig:schematic} also displays these BDT bounds as representative projections for two considered masses of the MuCol$\nu$ detector. Increasing the detector mass from $10~\textrm{kg}$ to $10~\textrm{ton}$ improves the projected bound by approximately an order of magnitude. While the number of expected BSM signal events scales as $N_{\textrm{sig}}\propto \lambda^2$ and increases roughly proportionally to the detector mass, the improvement in the bound is less than the naively expected factor of $10^{3/2}\approx 30$ relevant to the zero-background case. This is due to the increased difficulty of suppressing $\nu$-induced backgrounds in the larger detector, where the expected background rate after the BDT cut is $\mathcal{O}(100)$ events, compared to $\mathcal{O}(1)$ events in the smaller detector.

Furthermore, a detector with increased target mass may face greater challenges in muon identification and signal discrimination. This could limit the effective search region to the downstream section of the detector, in front of the spectrometer, where muon (mis)identification is more easily addressed. Consequently, the sensitivity difference between $10~\textrm{kg}$ and $10~\textrm{ton}$ detectors might be smaller than anticipated, further emphasizing the physics potential of the smaller experiment.

Nevertheless, these detectors can probe regions of the parameter space of the model inaccessible to other proposed searches. For comparison, we show projected bounds for DUNE~\cite{Kelly:2019wow,Berryman:2018ogk}, IC Gen-2~\cite{Esteban:2021tub}, and FLArE~\cite{Kelly:2021mcd}. Notably, even the small $10~\textrm{kg}$ detector can achieve comparable sensitivity to DUNE for $m_\phi\sim \textrm{a few} \times 100~\textrm{MeV}$, while the larger detector can extend the bounds in this mass range to $\lambda\lesssim 10^{-3}$. The proposed detectors also surpass the expected sensitivity of searches for invisible Higgs decays at the HL-LHC~\cite{deGouvea:2019qaz} and MuCol; see discussion below. This allows us to explore interesting dark matter targets for scalar masses between $m_\phi \sim 1~\textrm{MeV}$ and $20~\textrm{GeV}$, as discussed in \cref{sec:model}. Probing even heavier BSM particles could be possible with a detector operating at a MuCol energy of $10~\textrm{TeV}$.

\section{Invisible Higgs Decays}
\label{sec:Higgs}

The current constraints on heavy neutrino-philic scalars with mass $m_\phi\gtrsim m_K$ mainly come from measurements of rare weak boson decays~\cite{Berryman:2018ogk, deGouvea:2019qaz, Brdar:2020nbj}. This includes the decay $Z \to \nu \nu \phi$ which is constrained by the invisible Z branching fraction width measurement at LEP BR$(Z \!\to\! \text{inv.}) = 20 \pm 0.005 \%$~\cite{Electroweak:2003ram}; the decay $W \to \ell \nu \phi$ which is constrained by the $W \to \ell \nu$ branching fraction as measured by the LHC~\cite{ATLAS:2017rzl}; and the decay $h \to \nu \nu \phi$ (arising from the $\nu\nu\phi h$ term in the Lagrangian) which is constrained by the invisible Higgs branching fraction width measurement at the LHC BR$(h \!\to\! \text{inv.}) < 0.25$~\cite{CMS:2016dhk}.

A muon collider will likely not improve the measurement of the $Z$ and $W$ decay widths. However, it is expected to improve the measurement of the invisible Higgs decay width. The study performed in Ref.~\cite{Ruhdorfer:2023uea} found that a 3~TeV muon collider with luminosity of 2~ab$^{-1}$ could constrain the BSM contribution to $BR(h \to \text{inv.}) < 5 \cdot 10^{-3}$. This assumes that forward muon detectors with a coverage extending to $\eta \sim 5$ are installed on both sides, which would allow to measure the invariant mass of the invisibly decaying Higgs boson. 

We can recast this result to obtain a constraint on the neutrino-philic scalar parameter space. The Higgs decay width is given by 
\be
\Gamma (h \to \nu_\alpha \nu_\beta \phi) =
\frac{4 \lambda_{\alpha\beta}^2}{m_h v^2 (1+\delta_{\alpha\beta})} \int {\rm d} \Phi_3 
\left|{\cal M} \right|^2 
\ee
with expressions for the 3-body phase space $d\Phi_3$ and matrix element $\mathcal{M}$ provided in Ref.~\cite{deGouvea:2019qaz}. Analytically evaluating the integral, we obtain 
\be
\text{BR} &(h \!\to\!\nu_\alpha\nu_\beta\phi) = \frac{1}{\Gamma_h}\frac{ \lambda_{\alpha\beta}^2 m_h^3}{ 3\cdot 2^7 \pi^3 v^2 (1\!+\!\delta_{\alpha_\beta})}  \times \\
&\Big[(1\!-\!x)[x(x\!+\!10)\!+\!1] 
 +6x(1\!+\!x)\log(x) \Big]   .
\ee
with $x = m_\phi^2 / m_h^2$. For $m_{\phi} \ll m_h$ this simplifies to
\be
\text{BR} &(h\to\nu_\alpha\nu_\beta\phi) = \frac{1}{\Gamma_h}\frac{ \lambda_{\alpha\beta}^2 m_h^3}{ 3\cdot 2^7 \pi^3 v^2 (1\!+\!\delta_{\alpha_\beta})},  
\ee
which agrees with the result obtained in Ref.~\cite{Berryman:2018ogk}. Finally, let us note that both the neutrino and anti-neutrino channel contribute to the invisible Higgs decay width, $\text{BR}(h \!\to\! \text{inv.}) = \text{BR}(h \!\to\! \nu \nu \phi) + \text{BR}(h \!\to\! \bar\nu \bar\nu \phi)$.

Using the projected precision of the muon collider measurement $\text{BR}(h \!\to\! \text{inv.}) < 5 \cdot 10^{-3}$. We can estimate the projected sensitivity reach. For small $m_\phi$, the 3~TeV muon collider will be able to probe $\lambda > 0.086$. A full mass dependent sensitivity curve is shown all dash-dotted line in the right panel of \cref{fig:schematic}. We can see that it will surpass the sensitivity of LHC measurements. However, measurements using neutrino scattering at a muon collider are expected to be significantly more sensitive. 

\section{Conclusion}
\label{sec:conclusion}

Understanding the microscopic nature of the dark sector of the Universe remains the most pressing challenge in particle physics nowadays. Neutrinos have been recognized as a potential portal to this new physics, yet their tiny interaction rates make such endeavors particularly difficult. Powerful neutrino beams, ideally produced in laboratory settings, are needed to test this possibility thoroughly.

To this end, we have proposed to utilize a high-energy neutrino beam expected to be produced in the forward kinematic region of the future Muon Collider to search for a neutrinophilic scalar $\phi$, with a mass up to a few tens of GeV. Such scalars could mediate the interaction between dark matter and the Standard Model, providing particularly demanding thermal targets that will miss detection in traditional DM searches. Yet another compelling possibility is that DM is composed of an additional sterile neutrino with a mass of about $10~\textrm{keV}$ and couplings to the SM driven by the mixing with active neutrinos, which gain substantial self-interaction rates via $\phi$. This non-thermal DM target can also be indirectly probed by searching for the mediator $\phi$.

We have shown that a compact forward neutrino detector at the MuCol, i.e., the MuCol$\nu$ detector, with a mass of about $10~\textrm{kg}$, can probe these theory targets even within a year of operation. While the search for the $\phi$ scalar suffers from substantial neutrino-induced backgrounds, these can be suppressed by several orders of magnitude by observing an apparent lepton-flavor violation in the $\phi$ production in neutrino scatterings and by imposing further kinematical cuts to isolate signal events that we have discussed in detail.

The expected sensitivity can be further improved by employing a larger target material mass, and we have also considered discovery prospects for the $10~\textrm{ton}$ MuCol$\nu$ detector. Notably, this also represents the expected sensitivity of, e.g., a $1~\textrm{ton}$ detector operating for $10$ years. This ($1~\textrm{ton}$) detector mass corresponds to currently operating forward neutrino experiments at the LHC~\cite{FASER:2019dxq, FASER:2020gpr, FASER:2022hcn, Ahdida:2750060, SNDLHC:2022ihg}, while even larger detectors have been proposed for the future Forward Physics Facility at the LHC~\cite{Anchordoqui:2021ghd, Feng:2022inv, Adhikary:2024nlv}, which would pave the way for the search discussed in this study.

The positive muon production in high-energy muon neutrino scatterings can also be due to the neutrino trident production, however these events will typically be associated with a high-energy negative muon in the final state and will be rejected in our analysis. Dedicated studies of neutrino-trident scattering at high energies, suitable for the FPF era, have been performed in Refs.~\cite{Francener:2024wul,Altmannshofer:2024hqd,Bigaran:2024zxk} -- we expect similarly compelling work focused on MuCol neutrinos in the near future. This will provide a guaranteed physics case for the MuCol$\nu$ detector in the near future. Measurements using a MuCol neutrino source (and its very small flux uncertainties) will be highly complementary to those from LHC-produced neutrinos in better understanding the physics underlying neutrino trident scattering.

The forward neutrino beam at the MuCol is by far the largest among all high-energy colliders, offering an unprecedented sensitivity reach in the considered search. This further motivates the work on the proposed forward neutrino detector at the Muon Collider.

\section*{Acknowledgements}

We thank Max Fieg and Andrea Wulzer for sharing the neutrino flux obtained in Ref.~\cite{InternationalMuonCollider:2024jyv}.  
The work of F.K.~was supported by the Deutsche Forschungsgemeinschaft under Germany's Excellence Strategy -- EXC 2121 Quantum Universe -- 390833306. 
JA and ST are supported by the National Science Centre, Poland (research grant No. 2021/42/E/ST2/00031). KJK is supported in part by DOE Grant No. DE-SC0010813.

\bibliography{ref}

\end{document}